\documentclass[aps,pre,twocolumn,amsmath,amssymb,superscriptaddress]{revtex4-1}

\usepackage{graphicx}
\usepackage{cases}
\usepackage{dcolumn}
\usepackage[english]{babel}
\usepackage{bm}
\usepackage[dvipsnames]{xcolor}
\usepackage{letltxmacro}
\usepackage{xr-hyper}
\usepackage{hyperref}
\newcommand\numberthis{\addtocounter{equation}{1}\tag{\theequation}}
\LetLtxMacro{\ORIGselectlanguage}{\selectlanguage}
\makeatletter
\DeclareRobustCommand{\selectlanguage}[1]{%
  \@ifundefined{alias@\string#1}
    {\ORIGselectlanguage{#1}}
    {\begingroup\edef\x{\endgroup
       \noexpand\ORIGselectlanguage{\@nameuse{alias@#1}}}\x}%
}
\newcommand{\definelanguagealias}[2]{%
  \@namedef{alias@#1}{#2}%
}
\makeatother
\definelanguagealias{en}{english}

\newcommand{\DRD}{D^{\rm (RD)}}
\newcommand{\DYS}{D^{\rm (YS)}}
\newcommand{\sigmaeq}{\sigma_{\rm eq}}

\begin{document}
\title{Scaling in local to global condensation of wealth on sparse networks}
\author{Hyun Gyu Lee}
\email{supersting85@kias.re.kr}
\affiliation{School of Computational Sciences, Korea Institute for Advanced Study, Seoul 02455, Korea}
\author{Deok-Sun Lee}
 \email{deoksunlee@kias.re.kr}
\affiliation{School of Computational Sciences, Korea Institute for Advanced Study, Seoul 02455, Korea}
\affiliation{Center for AI and Natural Sciences, Korea Institute for Advanced Study, Seoul 02455, Korea}
\date{\today}

\begin{abstract}
The prevalence of wealth inequality propels us to characterize its origin and progression, via empirical and theoretical studies. The Yard-Sale(YS) model, in which a portion of the smaller wealth is transferred between two individuals, culminates in the concentration of almost all wealth to a single individual, while distributing rest of the wealth with a power-law of exponent one. By incorporating redistribution to the model, in which the transferred wealth is proportional to the sender's wealth, we show that such extreme inequality is suppressed if the frequency ratio of redistribution to the YS-type exchange exceeds the inverse of the population size. Studying our model on a sparsely-connected population, we find that the wealth inequality ceases to grow for a period, when local rich nodes can no longer acquire wealth from their broke nearest neighbors. Subsequently, inequality resumes growth due to the redistribution effect by allowing locally amassed wealth to move and coalesce. Analyzing the Langevin equations and the coalescing random walk on complex networks, we elucidate the scaling behaviors of wealth inequality in those multiple phases. These findings reveal the influence of network structure on wealth distribution, offering a novel perspective on wealth inequality.
\end{abstract}
\maketitle

\section{Introduction}
\label{sec:intro}

Wealth inequality may be attributed to numerous socioeconomic factors and their orchestration. Yet the universal power-law wealth and income distributions~\cite{pareto_1897,levy_power_1996, xavier_power_2009} imply a common mechanism at play, and the possibility of understanding how wealth inequality has arisen and how long it will persist ~\cite{piketty2017capital, chancel_piketty_2021, chancel2022world}. Given that individuals participating in a trade can undergo wealth transfer due to imperfect pricing, various wealth exchange models have been studied extensively, and their steady-state solutions, often available analytically, serve as plausible explanations for various  wealth distributions~\cite{ispolatov_wealth_1998, bouchaud_wealth_2000,  yakovenko_stats_money_2000, agent_based_models_2007,stats_mech_of_money_2009,cha_patterns_2010,barbier_urn_2017}. 

The Yard-Sale(YS) model ~\cite{chakraborti_statistical_2000,hayes_computing_2002} is remarkable as it generates an extreme wealth inequality from a seemingly fair (and thus realistic) exchange rule:  A fraction of the sender's and receiver's smaller wealth is transferred in each trade, and ultimately, almost all wealth is concentrated in a single individual while the remaining wealth is distributed by a power-law with exponent one across the rest of the population~\cite{boghosian_kinetics_2014}. The model's simplicity and yet the emergence of such stark inequality have attracted much attention~\cite{chakraborti_distributions_2002, Sinha_2003, moukarzel_wealth_2007, Moukarzel_2011,bustos_guajardo_yard_sale_2012, boghosian_kinetics_2014, boghosian_oligarchy_2017}. The Gini index is a Lyapunov functional, never decreasing with time, in the YS model~\cite{boghosian_oligarchy_2017} and in a class of models~\cite{Francisco-Cardoso:2023aa}. However, such global wealth condensation does not yet happen in the real world;  The present era may be shorter than the condensation time scale.

The degree of real-world wealth inequality has been changing with time~\cite{piketty2017capital,chancel_piketty_2021,chancel2022world}. In this light, the non-stationary, rather than stationary, state of a model may offer a better explanation of the reality. Also, by relaxing constraints like the fully-connected population often assumed in many studies and by considering multiple modes of wealth exchange, the YS model can become more realistic and reveal a richer set of insights. For example, transferring a fraction of the sender's wealth, occurring in donation, investment, or taxation and also called the {\it loser} rule~\cite{hayes_computing_2002,Cardoso:2020aa}, effectively redistributes wealth and suppresses wealth inequality~\cite{ispolatov_wealth_1998, bouchaud_wealth_2000, yakovenko_stats_money_2000, Sinha_2003} while there can be various ways to achieve redistribution~\cite{Cardoso:2020aa}. Investigating the non-stationary state of the generalized YS model~\cite{boghosian_kinetics_2014, boghosian_oligarchy_2017} which allows such redistribution (RD) mode of wealth transfer, as well as the YS-mode, between connected pairs in a structured population~\cite{bustos_guajardo_yard_sale_2012}, we identify new  factors influencing wealth inequality and provide a novel theoretical framework. 

We show that the extreme inequality of the original YS model in the long-time limit can be suppressed by increasing the ratio of the RD-mode transfers beyond the inverse of the population size. In the sparsely-connected population, inequality evolves with time through multiple phases, and we elucidate the underlying mechanisms. Initially, the inequality grows primarily driven by the YS-mode transfers before saturating over a period of time due to depleted wealth of the nearest neighbors of locally rich nodes. We call this stage {\it local condensate} phase. As time passes, the RD-mode transfer effectively thaws this frozen state, enabling further elevation of inequality via random walk and coalescence of locally-concentrated wealth. In these stages, wealth inequality exhibits scaling behaviors, which we show originate from the correlation between wealth and connectivity of individuals. This demonstrates the critical role of the structure of networks in shaping the wealth distribution. Finally, comparing with the empirical data, we discuss the implications of our findings.

\section{Model}
\label{sec:model}

\begin{figure}
\includegraphics[width=\columnwidth]{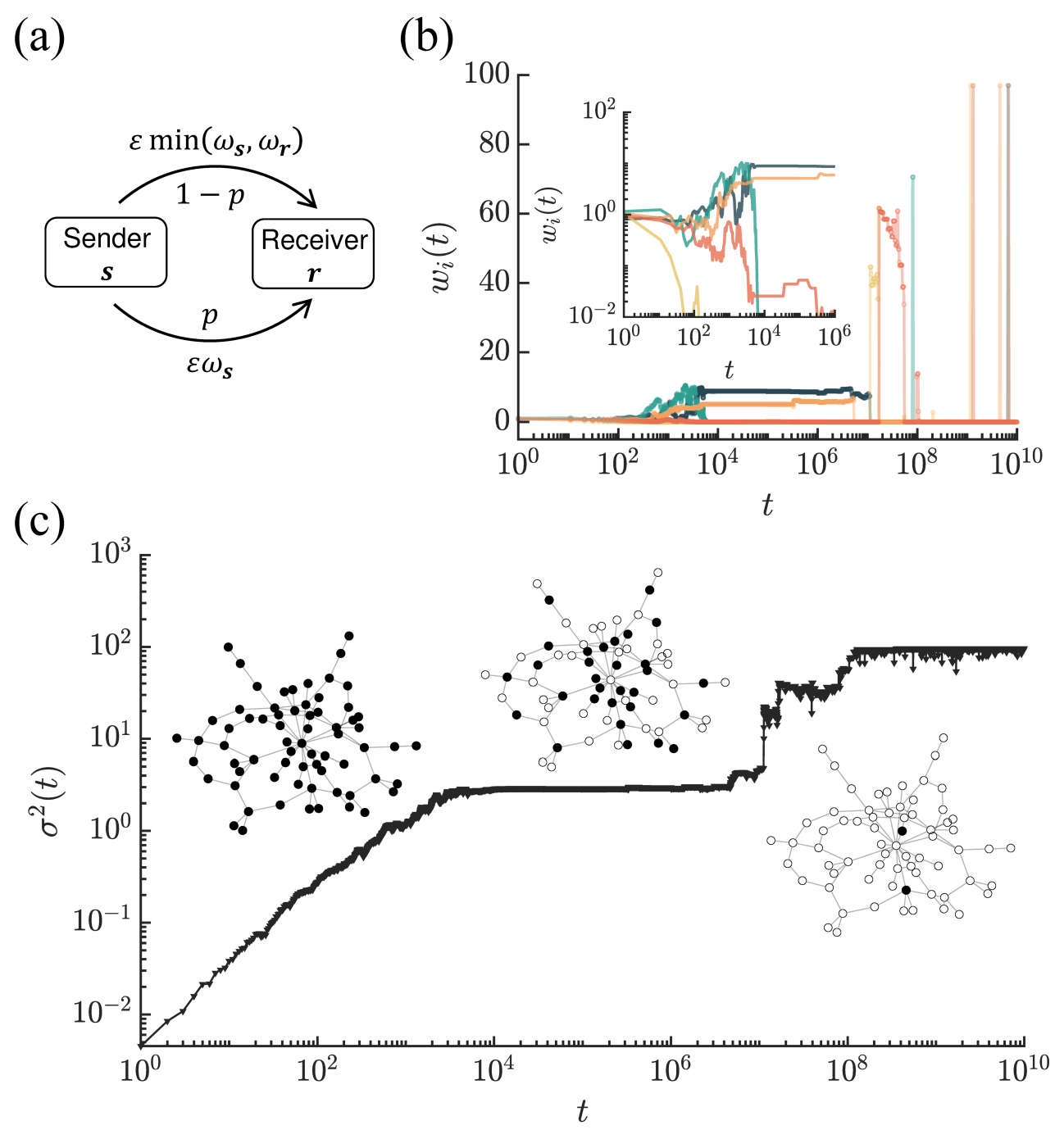}
\caption{Model and its Monte-Carlo simulation results. 
(a) Two modes of wealth transfers in Eq.~(\ref{eq:wealthtransfer}). 
(b) Time-evolution of individual wealth in a single run of simulation with $p=10^{-6}$ and $\varepsilon=0.05$ on a SF network of $N=97$ nodes, $L=200$ links 
 and the degree exponent $\gamma=2.5$. Inset: The same plots for $1 \le t \le 10^6$.
(c) Time-evolution of wealth variance from the same simulation. Three insets represent a part of the network 
at $t=10^2$, $10^5$, and $10^{10}$, respectively, with rich (poor) nodes colored black (white).}
\label{fig:gYS} 
\end{figure}

We consider a network of $N$ nodes (individuals) connected by $L$ undirected links (trade partnership) with  the adjacency matrix $A_{ij}=0, 1$. Each node $i$ has wealth $\omega_i(t)$ with $\omega_i(0)=1$ initially.  For every pair of connected nodes with rate $N/L$, the sender (`s') and the receiver (`r') are determined randomly and the sender sends an amount $\Delta \omega$ of wealth to the receiver [Fig.~\ref{fig:gYS} (a)], where $\Delta \omega$ is a fraction $\varepsilon$ of either the smaller wealth or the sender's wealth as 
\begin{equation}
\Delta \omega = 
\begin{cases}
\varepsilon \, \min \{\omega_{\rm s}, \omega_{\rm r} \} & \ {\rm with} \ {\rm prob.} \ 1-p  \ ({\rm YS \ mode}),\\
\varepsilon \, \omega_{\rm s} & \ {\rm with} \ {\rm prob.} \ p \ ({\rm RD\ mode}).
\end{cases}
\label{eq:wealthtransfer}
\end{equation}
Consequently their wealth change as $(\omega_{\rm s}, \omega_{\rm r})\to (\omega_{\rm s}-\Delta \omega, \omega_{\rm r}+\Delta \omega)$ while their sum is preserved. The mean wealth is fixed, i.e., $\overline{\omega}\equiv N^{-1}\sum_i \omega_i(t)=1$. We use $\overline{x}=N^{-1}\sum_{j=1}^N x_j$ to denote the spatial average. The parameter $p$ is the relative ratio of the RD-mode transfers. This is a network version of the model introduced in Refs.~\cite{boghosian_kinetics_2014, boghosian_oligarchy_2017}.  
For the underlying networks, we use  the complete graphs ($A_{ij}=1$ for all $i\neq j$) and the giant connected components of sparse scale-free (SF) networks~\cite{barabasi_scale_free_1999} constructed by the static model~\cite{goh_universal_2001,lee_evolution_2004}, which display power-law degree distributions $P_{\rm deg}(k) \equiv N^{-1}\sum_i \delta_{k_i,k} \sim k^{-\gamma}$ for large $k$ with the degree $k_i = \sum_j A_{ij}$ meaning the number of the nearest neighbors and $\gamma$ called the degree exponent, and have the mean degree $\overline{k}=2L/N$ finite. Note that for a unit time interval, every pair of nodes is selected ${2\over \overline{k}}$ times on the average for the transaction in Eq.~\eqref{eq:wealthtransfer}.

For a measure of wealth inequality we use the wealth variance 
$
\sigma^2(t)  \equiv \frac{1}{N} \sum_{i=1}^{N}(\omega_i(t)-\overline{\omega})^2,
$
the second cumulant of the wealth distribution $P(\omega,t)$. A single run of the model simulation with small $p$ readily reveals multiple phases in the time-evolution of wealth inequality. See Figs.~\ref{fig:gYS}(b) and \ref{fig:gYS}(c).
 i) In the early-time regime, for $t\lesssim 10^4$, individuals' wealth is made increasingly different from one another, so wealth inequality grows.
 ii) Then a frozen period follows ($10^4 \lesssim t \lesssim 10^7$), when $\omega_i$'s and  $\sigma^2$ hardly change with time. {\it Rich} nodes, with at least mean wealth ($\omega\geq 1$), are surrounded by the {\it poor} nearest neighbors having $\omega<1$.  With $p=0$, this local condensate phase becomes the equilibrium~\cite{bustos_guajardo_yard_sale_2012}. 
iii) For $10^7 \lesssim t \lesssim 10^8$, the wealth variance resumes growing and each locally-concentrated wealth switches its host  to one of its nearest neighbors repeatedly, appearing to perform a random walk, until it encounters another local wealth and they coalesce~\cite{github_link}. 
iv) In the late-time regime ($t\gtrsim10^8$), global condensation occurs; almost all wealth is concentrated onto a single node.  Yet its host changes with time  and $\sigma^2$ fluctuates though weakly. 

\section{Early and late-time regime: Analytic results}
\label{sec:analytic}

To understand these observations quantitatively and proceed, we construct the Langevin equation.  In the mean-field approximation, it provides analytic results  for the wealth variance and distribution in the early and late-time regime.
\subsection{Langevin equation}
\label{sec:langevin}

In our model, at each discrete time step $\tau$,  a pair of connected nodes ($i, j$) are selected among $L$ pairs and performs the YS-mode transfer with probability $1-p$ or the RD-mode transfer with probability $p$. The node $i$ or $j$ becomes the sender with equal probability.  Therefore the expected change of the wealth $\omega_i$ of a node $i$ for the period from $\tau_0$ to $\tau_0 + \Delta\tau$ is given by
\begin{align*}  
    \Delta \omega_i  &\equiv \langle \{\omega_i(\tau_0 + \Delta \tau) - \omega_i(\tau_0)\}\rangle  \\
    & = \varepsilon \sum_{j=1}^{N} A_{ij} \biggl[p\frac{\Delta \tau}{L} \frac{\omega_j-\omega_i}{2} + \sqrt{p\frac{\Delta \tau}{L}} \frac{\omega_i+\omega_j}{2} \eta_{ij} \\
     &+ \sqrt{(1-p)\frac{\Delta \tau}{L}} \min{(\omega_i,\omega_j)} \xi_{ij} \biggr] \numberthis
\end{align*}
with the continuous random variables $\eta_{ij}$ and $\xi_{ij}$ satisfying $\langle \eta_{ij} \rangle = \langle \xi_{ij} \rangle = 0$, $\langle \eta_{ij}^2 \rangle = \langle \xi_{ij}^2 \rangle = 1$, $\eta_{ij} = -\eta_{ji}$, and $\xi_{ij} = -\xi_{ji}$. We use $\langle \cdots\rangle$ to represent the ensemble average. Introducing (continuous) time $t\equiv \frac{\tau}{N}$ and taking large-$N$ limit, we find
\begin{align*}
    d \omega_i = & -\frac{\varepsilon p}{\overline{k}} \sum_{j}L_{ij}\omega_j dt  
    + \varepsilon \sqrt{\frac{2p}{\overline{k}}} \sum_{j}A_{ij} \frac{\omega_i+\omega_j}{2} dC_{ij} \\
    & + \varepsilon \sqrt{\frac{2(1-p)}{\overline{k}}} \sum_{j}A_{ij} \min{(\omega_i,\omega_j)} dB_{ij},   \numberthis
\end{align*}
where $L_{ij}\equiv k_i \delta_{ij} - A_{ij}$ is  the Laplacian, $dB_{ij}$ and $dC_{ij}$ are the Wiener processes satisfying $\langle dB_{ij} \rangle = \langle dC_{ij} \rangle = 0$, $\langle dB_{ij}^2 \rangle = \langle dC_{ij}^2 \rangle = dt$, $dB_{ij} = -dB_{ji}$, and $dC_{ij} = -dC_{ji}$. To fulfill the condition $\sum_i d\omega_i = 0$, we adopted It$\hat{o}$'s scheme. 

To proceed, we approximate the sum of the YS- and RD-mode transfers with all the nearest neighbors as
 $ \sqrt{\frac{1}{\overline{k}}} \sum_j A_{ij} \frac{\omega_i + \omega_j}{2} dC_{ij} \simeq \sqrt{\DRD(\omega_i)} dX_i$ and 
$ \sqrt{\frac{1}{\overline{k}}} \sum_j A_{ij} \min{(\omega_i,\omega_j)} dB_{ij} \simeq \sqrt{\DYS(\omega_i)} dY_i$,  where $dX_{i}$ and $dY_{i}$ are the Wiener processes satisfying $\langle dX_{i} \rangle = \langle dY_{i} \rangle = 0$, $\langle dX_{i}^2 \rangle = \langle dY_{i}^2 \rangle = dt$. The coefficients $\DRD(\omega)= \left\langle \overline{k}^{-1}\sum_{j=1}^{N} A_{ij} \{(\omega_i + \omega)/2\}^2 \right\rangle$ and $\DYS(\omega)=\left\langle \overline{k}^{-1} \sum_{j=1}^{N} A_{ij}  \min\{\omega,\omega_j\}^2 \right\rangle$ are the mean square of the transferred wealth for node $i$ under the RD and YS mode and  given approximately by
\begin{align}
\DRD(\omega)&\simeq \langle \overline{\{(\omega_i + \omega)/2\}^2}\rangle \simeq 
\begin{cases}
{\langle\overline{\omega^2}\rangle\over 4} & {\rm for} \ \omega\to 0,\\
{\omega^2 \over 4} & {\rm for} \ \omega\to \infty,
\end{cases}
\cr
\DYS(\omega)&\simeq  \langle \overline{\min\{\omega_i, \omega\}^2}\rangle \simeq 
\begin{cases}
\omega^2 & {\rm for} \ \omega\to 0,\\
\langle\overline{\omega^2} \rangle& {\rm for} \ \omega\to \infty.
\end{cases}
\label{eq:D}
\end{align}
Also note that $\DRD(\omega)\simeq 1$ and $\DYS(\omega)\simeq 1$ for $\omega$ close to the initial value $1$ in the early-time regime when all nodes have wealth close to $1$.  Finally, the Langevin equation for the wealth of a node  is represented as 
\begin{align*}
d\omega_i= & -\frac{\varepsilon \, p}{\overline{k}} \sum_{j}L_{ij}\omega_j dt  + \varepsilon \sqrt{2p\, \DRD(\omega_i)} \, dX_i \\
&+ \varepsilon \sqrt{2(1-p)\, \DYS(\omega_i)} \, dY_i,  \numberthis
\label{eq:Langevin}
\end{align*}
where $dX_i$ and $dY_i$ are the Wiener processes with mean $0$ and variance $dt$ representing the stochasticity of whether to send or receive wealth, and the coefficients $\DRD(\omega_i)$ and $\DYS(\omega_i)$ are given in Eq.~\eqref{eq:D}.  The first and second terms represent the deterministic and stochastic changes by the RD-mode transfers, and the third one from the YS transfers. 

\subsection{Wealth variance }
\label{sec:variance}

\begin{figure*}
\includegraphics[width=2\columnwidth]{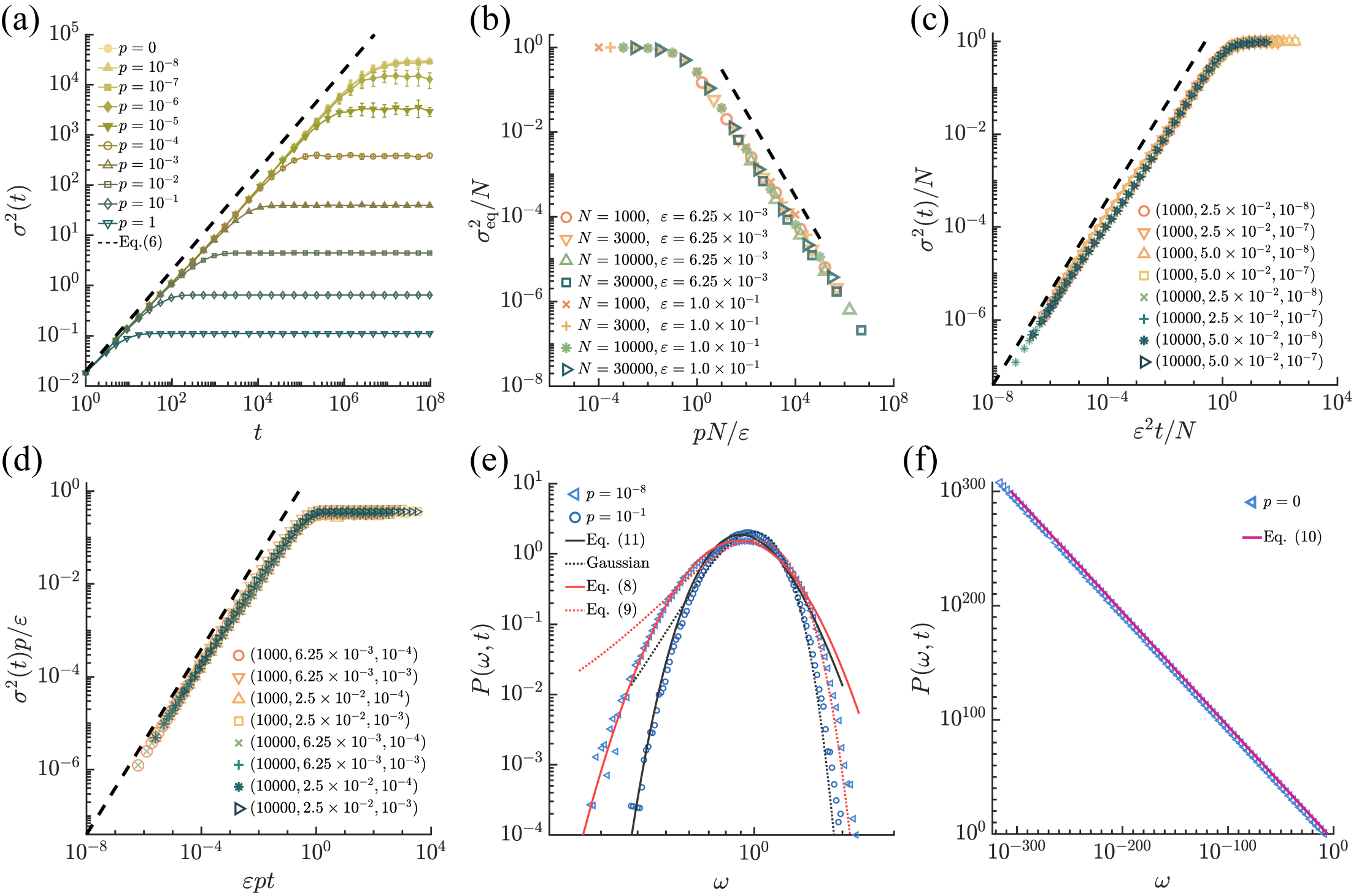}
\caption{Wealth variance and distribution on the complete graphs. 
(a) Wealth variance for $\varepsilon=0.1$ and different $p$'s   with $N=10^5$ averaged over 20 realization.
(b) Data collapse of the rescaled wealth variance in the equilibrium state for 
different $p$'s,  
$\varepsilon$'s and $N$'s. 
The dashed line has slope $-1$. 
(c) Plots of $\sigma^2/N$ versus $\varepsilon^2 t/N$ for the parameters satisfying $p< 0.1 p_*$ with $p_*\equiv {\varepsilon \over N}$. Legends represent $(N, \varepsilon, p)$. 
(d) Plot of $\sigma^2 p/\varepsilon$ versus $\varepsilon p t$ for the parameters satisfying $p > p_*$.
(e) Wealth distributions with different $p$'s for $N=10^5$ and $\varepsilon=0.00625$ at time $t=10^3$. Lines represent  analytic predictions.  `Gaussian' denotes $P(\omega,t) = \frac{1}{\sqrt{2\pi\sigma^2(t)}} e^{-\frac{(\omega-1)^2}{2\sigma^2(t)}}$.
(f) Wealth distribution with $p=0$ for $N=10^5$ and $\varepsilon=0.1$ at time $t=10^8$.
}
\label{fig:cg}
\end{figure*}

In the early-time regime, $\omega_i$'s remain close to $1$, and thus $\DYS(\omega_i)  \simeq \DRD(\omega_i)  \simeq 1$. Approximately we have $d\omega_i \simeq \sqrt{2}\varepsilon \, dY_i$ and $d\omega_i \simeq \sqrt{2}\varepsilon \, dX_i$ for small and large $p$, respectively, leading to 
\begin{equation}
\sigma^2(t) \simeq 2\varepsilon^2 t.
\label{eq:sigma2early}
\end{equation}
 See Figs.~\ref{fig:cg}(a). This linear growth cannot continue indefinitely for finite $N$ but $\sigma^2$ eventually saturates. The equilibrium value varies with $p$ as [Fig.~\ref{fig:cg}(b)]
\begin{align}
  \sigma^2_{\rm eq} \equiv \lim_{t\to\infty} \sigma^2(t) \sim 
  \begin{cases}
      N & \text{for \quad $p \ll p_*\equiv\frac{\varepsilon}{N}$,} \\
      \frac{\varepsilon}{p} & \text{for \quad $p \gg p_*$}.
    \end{cases} \numberthis
\label{eq:sigma2eq}
\end{align}
In the limit $N\to\infty$, the condensation is suppressed with any finite $p$, which has been shown in case of $\epsilon=1$ in Ref.~\cite{e25071105}. The critical ratio $p_*$ and Eq.~(\ref{eq:sigma2eq}) are obtained as follows. 
For $p=0$ or small $p$, like the original YS model, almost all wealth is concentrated in a single node in the long-time limit~\cite{boghosian_kinetics_2014} yielding $\sigmaeq^2\simeq N$. For $p$ relatively large,  let us take the mean-field approximation $\sum_{j}L_{ij}\omega_j \simeq \overline{k}(\omega_i - 1)$, which becomes exact in the complete graph. In equilibrium, the fluctuation driven by the YS-mode transfers $(\delta \omega)_{\rm YS} \sim \varepsilon \sqrt{2t_{\rm eq}D_{\rm YS}}$ is balanced by the redistribution $(\delta \omega)_{\rm RD} \sim \varepsilon p (\delta \omega)_{\rm YS} t_{\rm eq}$ in a time interval $t_{\rm eq}$, which allows us to estimate $t_{\rm eq} \sim \frac{1}{\varepsilon p}$ and  $\sigmaeq^2  \sim (\delta \omega)_{\rm YS}^2 \sim (\delta \omega)_{\rm RD}^2 \sim \frac{\varepsilon}{p}$. It  is the fluid phase of wealth. The diverging and finite $\sigma_{\rm eq}^2$'s become comparable at $p_* \equiv {\varepsilon\over N}$.  Simulations on the complete graphs support Eq.~(\ref{eq:sigma2eq}) [Fig.~\ref{fig:cg}(b)].

Given Eqs.~\eqref{eq:sigma2early} and \eqref{eq:sigma2eq}, the wealth variance can be represented as $\sigma^2(t) = \sigma_{\rm eq}^2 \Phi\left({t\over t_{\rm eq}}\right)$ with a function $\Phi(x)$ behaving as  $\Phi(x)\simeq x$ for $x\ll 1$ and $\Phi(x)\simeq 1$ for $x\gg 1$. 
The equilibration time  $t_{\rm eq}$ is given by $t_{\rm eq} \sim \frac{N}{\varepsilon^2}$ for $p \ll p_*$, and $t_{\rm eq} \sim \frac{1}{\varepsilon p}$ for $p \gg p_*$.  This is confirmed numerically by the collapse of the rescaled data in Fig.~\ref{fig:cg} (c) and (d). 

\subsection{Wealth distribution}
\label{sec:distribution}

The wealth distribution $P(\omega,t)$ is obtained, though partially, by the mean-field approach. With $p\ll p_*$,  Eq.~\eqref{eq:Langevin} can be approximated as $d\omega_i \simeq \varepsilon \sqrt{2 \DYS (\omega_i)} dY_i$ and the Fokker-Planck (FP) equation is given by $\partial P/\partial t \simeq \varepsilon^2 (\partial^2/\partial \omega^2) \{\DYS(\omega)P\}$. Recalling $\DYS(\omega) \simeq \omega^2$ for $\omega\ll 1$ and $D_{\rm YS}(\omega)\simeq 1$ for $\omega\gg 1$ in the early-time regime,  we find 
\begin{equation}
P(\omega,t)\simeq  {e^{-{t\varepsilon^2\over 4}} \over \sqrt{4\pi\varepsilon^2t} \, \omega^{3/2}} \exp \left[-\frac{(\log{\omega})^2}{4\varepsilon^2t} \right]
\label{eq:Plognormal}
\end{equation}
for $\omega\ll 1$ and 
\begin{equation}
P(\omega,t)\simeq  {1\over \sqrt{4\pi \varepsilon^2 t}} e^{-{(\omega-1)^2\over 4\varepsilon^2 t}} 
\label{eq:Pgaussian}
\end{equation}
for $\omega\gg 1$, respectively, in agreement with the simulation results [Fig.~\ref{fig:cg}(e)]. Note that the width of these distributions is commonly given by  $\langle \omega^2\rangle-1 \simeq 2\varepsilon^2 t$ for $\varepsilon^2 t\ll 1$ in agreement with Eq.~(\ref{eq:sigma2early}). 

In the long-time limit,  a single node occupies almost all wealth, slightly less than $N$. The  rest of the wealth, much less than $N$,  is distributed over the remaining $N-1$ nodes by a  power-law distribution 
\begin{equation}
P(\omega,t)\simeq {1\over \varepsilon^2 \,\omega\, t},
\label{eq:pwp0}
\end{equation}
which is supported numerically in Fig.~\ref{fig:cg}(f).
It is obtained by solving the Boltzmann equation investigated in Ref.~\cite{boghosian_kinetics_2014} and also presented in Appendix~\ref{seca:boltzmann}.

With $p\gg p_*$,  one can approximate Eq.~(\ref{eq:Langevin}) as $d\omega \simeq -\varepsilon p (\omega-1) dt + \varepsilon\sqrt{2(1-p)} \omega dY$ for small $\omega$ under the assumption $k_i/\overline{k}\simeq 1$. This Langevin equation including a multiplicative noise has been studied in~\cite{bouchaud_wealth_2000}  and also for the stationary state of the YS model with redistribution~\cite{boghosian_kinetics_2014, boghosian_oligarchy_2017}. The FP equation  is given by
$\partial P / \partial t = (\partial/\partial \omega)\{\varepsilon p (\omega - 1) P \} 
    + (\partial^2/\partial \omega^2) \Bigl\{\varepsilon^2 (1-p)\omega^2 P \Bigr\}$ in the small-$\omega$ region, and  one can see that the stationary-state solution is the inverse gamma distribution
\begin{eqnarray}
    P_{\rm eq}(\omega)\simeq \frac{\mu^{\mu+1}}{\Gamma(\mu+1)} 
    \omega^{-2-\mu}e^{-\frac{\mu}{\omega}},
    \label{eq:Pinversegamma}
\end{eqnarray}
where $\mu\equiv\frac{p}{\varepsilon(1-p)}\simeq \frac{p}{\varepsilon}$ and the width is $\langle\omega^2\rangle -1=\frac{1}{\mu-1}\simeq \frac{\varepsilon}{p}$ for large $\mu$. Note that it is a power-law $P_{\rm eq}(\omega)\sim \omega^{-2-{p\over \varepsilon}}$ for large $\omega$~\cite{bouchaud_wealth_2000} while Eq.~(\ref{eq:Pinversegamma}) describes only the small-$\omega$ behavior of our model.

\section{Scaling in the intermediate-time regime: Network effects}
\label{sec:intermediate}

\begin{figure}
\includegraphics[width=\columnwidth]{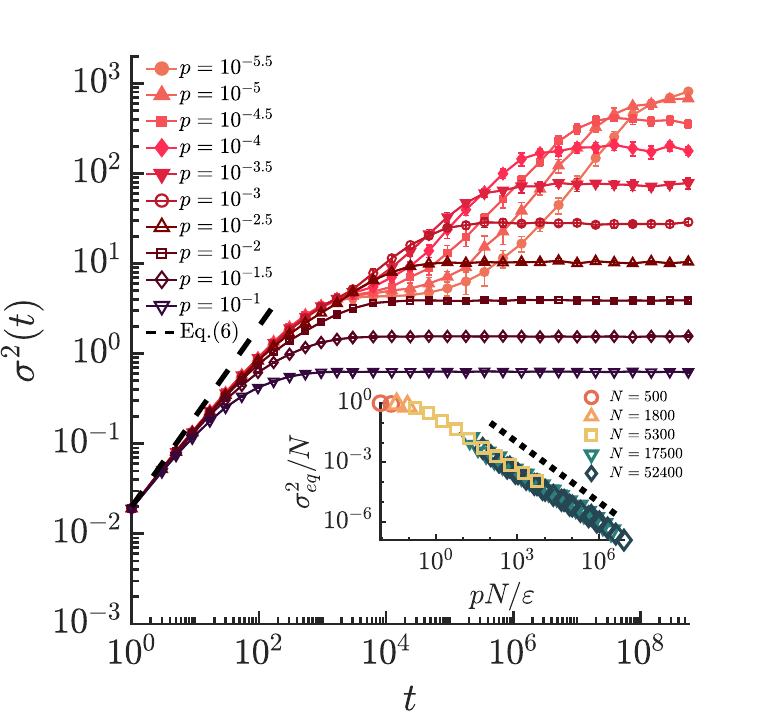}
\caption{Wealth variance with different $p$'s for $\varepsilon=0.1$  on SF networks of $N=887\pm13$, $\overline{k}=4.5$, and the degree exponent $\gamma=2.5$, averaged over 100 realizations. Inset: data collapse of the rescaled wealth variance in the equilibrium state. }
\label{fig:sfintro}
\end{figure}

For sparse networks, the early- and late-time behaviors remain similar to those on the complete graphs as studied in Sec.~\ref{sec:analytic}.  See Fig.~\ref{fig:sfintro}.  However, in the intermediate-time regime,  
the wealth variance ceases to grow but remains fixed for a period and then resumes growth.  These novel phases emerging on sparse networks are studied in this section, setting the ratio of RD transfers to be small.

\begin{figure*}
\includegraphics[width=2\columnwidth]{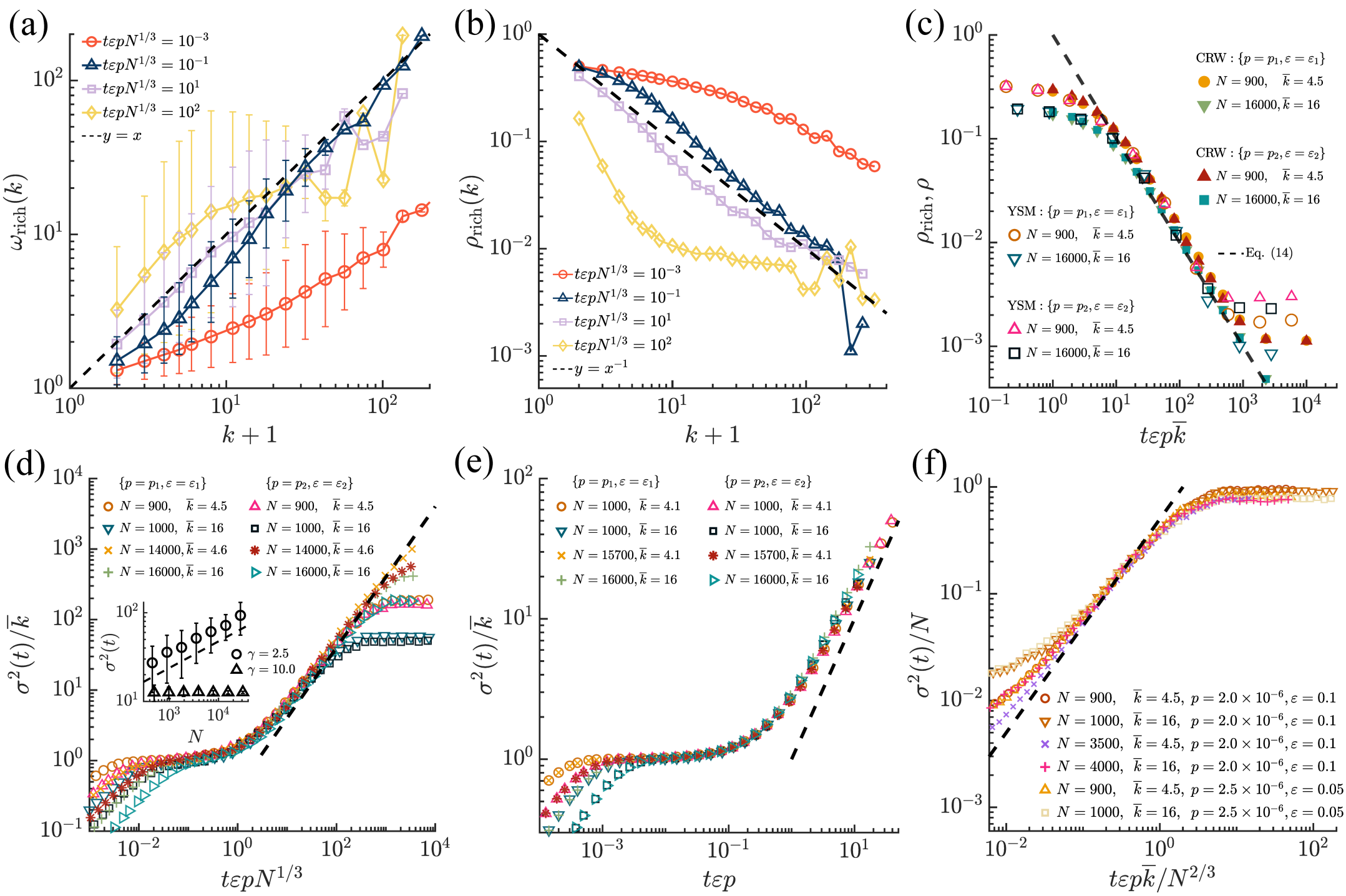}
\caption{Scaling of wealth variance on SF networks. 
(a) Wealth of rich nodes versus degree plus one at different times for $p=2.5\times 10^{-6}$ and $\varepsilon=0.05$  with $N=27989\pm27$, $\overline{k}=4.5$, and $\gamma=2.5$. 
(b) Plots of the probability $\rho_{\rm rich}(k)$ of a node of degree $k$ to be rich. 
(c) Time-decay of the rich-node density  $\rho_{\rm rich}$ in the generalized YS model compared with the random-walker density $\rho$ in the CRW  with $(p_1,\varepsilon_1) = (2.0\times 10^{-6}, 0.1)$ and  $(p_2,\varepsilon_2) = (2.5\times 10^{-6}, 0.05)$  for $\gamma=2.5$ and different $N$'s and $\overline{k}$'s. See Appendix~\ref{seca:CRW}. The dashed line is Eq.~(\ref{eq:rho}).
(d) Data collapse of the rescaled wealth variance as predicted in Eq.~(\ref{eq:sigma2relax}) with
$\{(p_1,\varepsilon_1), (p_2,\varepsilon_2)\}$ as in (c) for $\gamma=2.5$ and different $N$'s and $\overline{k}$'s.  Inset: Wealth variance versus $N$ at fixed time $t=1.25/\varepsilon p$ with $p=2.5\times 10^{-6}$ and $\varepsilon=0.05$ for two different $\gamma$'s. The dashed line has slope $1⁄3$. 
(e) The same plots as in (e) for $\gamma=10$. The dashed lines in (d) and (e) have slope $1$.
(f) Data collapse of the rescaled wealth variance  exhibiting a crossover from the relaxation phase to the global condensate phase around $t_{\rm eq} = {N^{2/3}\over \varepsilon p \overline{k}}$ with $p<0.1 p_*$ on the SF networks with  $\gamma=2.5$.
}
\label{fig:sf} 
\end{figure*}

\subsection{Local condensate phase} 
\label{sec:lc}

In the local condensate phase for $t_{\rm lc} \lesssim t \lesssim t_{\rm rel}$, the wealth variance is fixed, which is related to the sparse connection of the underlying network. Each rich node $i$ is found to have taken nearly all the wealth of its nearest neighbors, possessing $\omega_{i; {\rm rich}}  \simeq k_i+1$ including its own [Fig.~\ref{fig:sf}(a)]. Hub nodes thus possess more as long as they are rich. Yet the probability of a node to be rich decreases with its degree as $\rho_{i;{\rm rich}}\simeq {1\over k_i+1}$ [Fig.~\ref{fig:sf}(b)], for a node and its neighbor(s) are equally likely to be rich under the YS-mode transfer.

Introducing the wealth $\omega_{\rm rich}(k)$ of a rich node of degree $k$  and the probability $\rho_{\rm rich}(k)$ of a node of degree $k$ to be rich and approximating the wealth of a poor node to be zero, one can represent the wealth variance as
\begin{equation}
\sigma^2 \simeq \sum_k P_{\rm deg}(k) \left[ \rho_{\rm rich}(k) \left\{ \omega_{\rm rich}(k) - 1 \right\}^2 + 1-\rho_{\rm rich}(k)\right].
\label{eq:sigma2decomp}
\end{equation} 
Using $\rho_{\rm rich}(k)\simeq {1\over k+1}$ and $\omega_{\rm rich}\simeq k+1$, we obtain
\begin{equation}
\sigma^2\simeq \overline{k},
\label{eq:sigma2lc}
\end{equation}
as supported by Figs.~\ref{fig:sf}(d) and (e).
The onset time of local condensation $t_{\rm lc} \sim \frac{\overline{k}}{\varepsilon^2}$, at which Eq.~(\ref{eq:sigma2early}) crossovers to Eq.~(\ref{eq:sigma2lc}), can be rationalized  by considering that it takes time $\varepsilon^{-2}$ for a node to take the wealth of a neighbor by the YS-mode transfers, and it has  $\overline{k}$ such neighbors on the average. Local condensation is terminated at $t_{\rm rel}$, when the RD-mode transfers begin to redistribute significantly the wealth of the local rich nodes to their poor neighbors. 

\subsection{Relaxation phase}
\label{sec:relaxation}

 On the time scale  longer than $t_{\rm rel}$, the RD transfers occur frequently  enough to redistribute the locally-concentrated wealth  to a neighboring node. Also one of the two wealths on neighboring nodes can absorb the other~\cite{github_link}. Such coalescence of wealth drives wealth to a single or a few nodes until the stationary state is reached, and we call this period the {\it relaxation} phase ($t_{\rm rel} \lesssim t\lesssim t_{\rm eq}$). 
 
The wealth variance  exhibits remarkable scaling behaviors [Fig.~\ref{fig:sf}]. The dynamics of wealth in this phase can be understood by studying the coalescing random walk (CRW) on complex networks~\cite{catanzaro_diffusion_2005,park_branching_2020}, which allows us to evaluate $\rho_{\rm rich}(k)$ and $\omega_{\rm rich}(k)$ and use them in  Eq.~(\ref{eq:sigma2decomp}) to obtain $\sigma^2(t)$. In the CRW suited for our model, the following occurs for every link with rate $\lambda$: 
i) if the link is occupied by a walker (local wealth)  at one end node and empty at the other end, the walker moves to the latter, 
ii) if both ends are occupied by walkers, they coalesce leaving one walker at either end, and 
iii) if both ends are empty, nothing happens over the link. One can show that the time-decrease of the walker density is proportional to the square of the density and obtain the solution $\rho\simeq  {1\over \lambda \overline{k} t}$ as detailed in Appendix~\ref{seca:CRW}. The jump rate $\lambda$ is governed by the rate of RD transfers and given by $\lambda_{\rm gYS}\sim \varepsilon p$. Therefore the fraction of rich nodes $\rho_{\rm rich}$ in our model is given by
\begin{equation}
   \rho_{\rm rich} \simeq {1\over \varepsilon p \overline{k} t} 
\label{eq:rho}
\end{equation}
for large $t$. It is confirmed in simulations [Figs.~\ref{fig:sf}(c)]. 

As random movement and coalescence of local wealth proceeds,  the probability  to be rich $\rho_{\rm rich}(k)$ loses its dependence on degree $k$ [Fig.~\ref{fig:sf}(b)]. The wealth of a node remains proportional to its degree [Fig.~\ref{fig:sf}(a)] with the proportional coefficient increasing as the number of rich nodes decreases. Assuming $\omega_{\rm rich}(k) \simeq c\, k$ with $c$ a coefficient and $\rho_{\rm rich}(k)\simeq \rho_{\rm rich}$ in Eq.~(\ref{eq:rho}), one can use the unit-mean condition $\overline{\omega}\simeq \sum_k P_{\rm deg}(k) \rho_{\rm rich}(k)\,  c\, k\simeq \rho_{\rm rich}\,c\, \overline{k}=1$ to obtain
$c \simeq \frac{1}{\rho_{\rm rich}\overline{k}}$.  Using these results in Eq.~(\ref{eq:sigma2decomp}), we find 
\begin{equation}
    \sigma^2(t)\simeq \rho_{\rm rich}\, c^2 \, \overline{k^2}
    \sim
    \begin{cases}
    \varepsilon p \overline{k} N^{3-\gamma \over \gamma-1} t & {\rm for} \ 2<\gamma<3,\\
    \varepsilon p \overline{k} t & {\rm for} \ \gamma>3,
    \end{cases}
     \label{eq:sigma2relax}
\end{equation}
where we used  ${\overline{k^n}\over \overline{k}^n} \sim  \max\{1, N^{n-\gamma+1\over \gamma-1}\}$ 
for the static-model SF networks~\cite{lee_evolution_2004}. The data collapses of the scaled plots for different $\varepsilon, p, \overline{k}$ and $N$ in Fig.~\ref{fig:sf}(d) and \ref{fig:sf}(e)  confirm these scaling behaviors for $2<\gamma<3$ and $\gamma>3$, respectively.

\section{Multiple crossovers in wealth inequality}
\label{sec:multiple}

The results that we have obtained in the previous sections provide an overview of the multiple phases in the development of wealth inequality on sparse networks, depending on the ratio $p$ of the RD-mode transfers and the degree exponent $\gamma$.  From Eqs.~(\ref{eq:sigma2early}), (\ref{eq:sigma2eq}), (\ref{eq:sigma2lc}), and (\ref{eq:sigma2relax}), we can expect the wealth variance to behave approximately as 
\begin{equation}
\sigma^2(t)\sim 
\begin{cases}
\varepsilon^2 t & {\rm for} \ t\ll t_{\rm lc},\\
\overline{k} & {\rm for} \ t_{\rm lc}\ll t\ll t_{\rm rel},\\
\overline{k}\varepsilon p t \max\{1, N^{3-\gamma\over \gamma-1}\} & {\rm for} \ t_{\rm rel}\ll t \ll t_{\rm eq},\\
\min\{N, {\varepsilon\over p}\} & {\rm for} \ t\gg t_{\rm eq}.
\end{cases}
\label{eq:sigma2all}
\end{equation}
The scaled plots in Fig. ~\ref{fig:sf} (d) and (e) confirm the crossover around $t_{\rm rel}$ for $p\ll p_*$. The crossover around the equilibration time $t_{\rm eq}$ is also shown in the data collapse of the scaled plots in Fig.~\ref{fig:sf}(f). 

The crossover time scales can be estimated by comparing the behaviors of $\sigma^2(t)$ in adjacent time regimes in Eq.~\eqref{eq:sigma2all}
and are given by
\begin{equation}
t_{\rm lc} = {\overline{k}\over \varepsilon^2},
\label{eq:tlc}
\end{equation}
\begin{align}
t_{\rm rel} &={1\over \varepsilon p  \max\{1, N^{3-\gamma\over \gamma-1}\}} \cr 
&=
\begin{cases}
t_{\rm rel}^{\rm (het)}={1\over \varepsilon p N^{3-\gamma\over \gamma-1}} & {\rm for} \ 2<\gamma<3,\\
t_{\rm eq}^{\rm (hom)}={1\over \varepsilon p} & {\rm for} \ \gamma>3,
\end{cases}
\label{eq:trel}
\end{align}
\begin{align}
t_{\rm eq} &= {\min\{N, {\varepsilon\over p}\}  \over \overline{k}\varepsilon p t \max\{1, N^{3-\gamma\over \gamma-1}\}}\cr 
&=
\begin{cases}
t_{\rm eq,C}^{\rm (het)}={N^{2(\gamma-2)\over \gamma-1} \over \varepsilon p \overline{k}} & {\rm for} \ p\ll p_*, \ 2<\gamma<3,\\
t_{\rm eq,F}^{\rm (het)}={1\over p^2 \overline{k} N^{3-\gamma\over \gamma-1}} &  {\rm for} \ p\gg p_*, \ 2<\gamma<3,\\
t_{\rm eq,C}^{\rm (hom)}={N\over \varepsilon p \overline{k}} & {\rm for} \ p\ll p_*, \ \gamma>3,\\
t_{\rm eq,F}^{\rm (hom)}={1\over p^2 \overline{k}} & {\rm for} \ p\gg p_*, \ \gamma>3,\\
\end{cases}
\label{eq:teq}
\end{align}
and the critical ratio $p_*$ distinguishing the global condensate phase and fluid phase is given by $p_*\equiv {\varepsilon \over N}$.  We remark that on the complete graphs ($\overline{k}=N-1$), the local condensate phase in Eq.~(\ref{eq:sigma2all}) is identical to the global condensate phase $\sigma^2\simeq N$ and therefore $t_{\rm lc}$ in Eq.~(\ref{eq:tlc}) should be considered as the equilibration time. 

\begin{figure}
\includegraphics[width=\columnwidth]{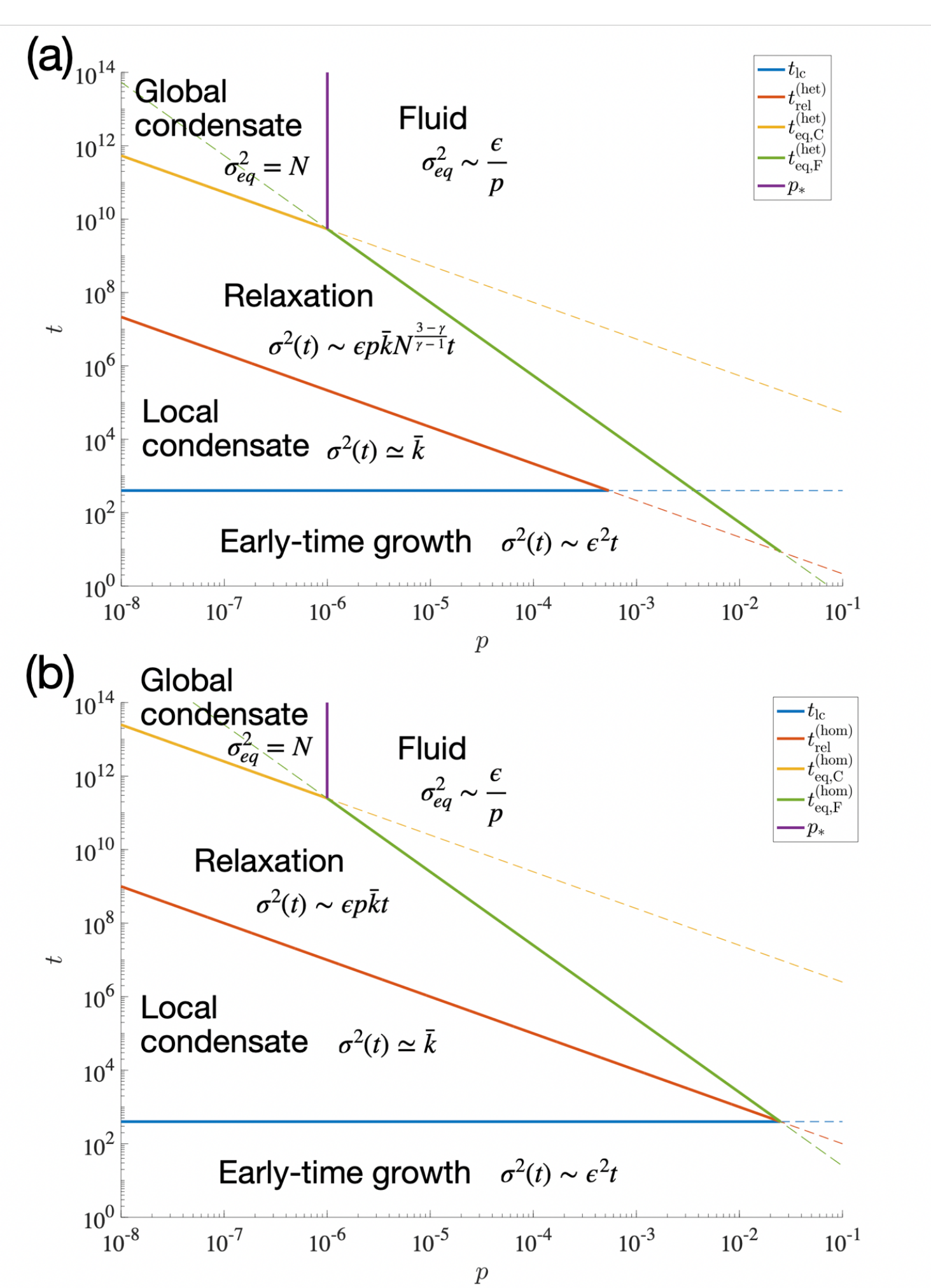}
\caption{Multiple phases of wealth inequality in the ($p,t$) plane for the generalized YS model with $\varepsilon=0.1$ on SF networks of $N=10^5, \overline{k}=4$ and (a) $\gamma=2.5$ and (b) $\gamma=10$. See Eq.~\eqref{eq:sigma2all}. The phase boundaries are given by the crossover time scales  in Eqs.~\eqref{eq:tlc}, \eqref{eq:trel}, and \eqref{eq:teq}.
}
\label{fig:PDall}
\end{figure}
From Eqs.~\eqref{eq:sigma2all}, \eqref{eq:tlc}, \eqref{eq:trel}, and \eqref{eq:teq}, we can obtain the diagram of different phases as in Fig.~\ref{fig:PDall}(a) and \ref{fig:PDall}(b) for $2<\gamma<3$ and $\gamma>3$, respectively.


In these results, the influence of the network structure is significant: A large and heterogeneous network  (small $\gamma$) facilitates wealth condensation by large wealth inequality and small $t_{\rm rel}$ and $t_{\rm eq}$ as seen in Eqs.~\eqref{eq:sigma2all} to \eqref{eq:teq}. Moreover, the correlation of  wealth and node degree leads the wealth distribution to share the similar asymptotic behaviors with the degree distribution as will be detailed in the below. It is also remarkable that the RD transfers enable wealth inequality to resume growth but eventually decrease the stationary-state wealth inequality. The increase of the fraction $\varepsilon$ not only speeds up the whole process, as shown  in Eq.~(\ref{eq:Langevin}), but also increases stochasticity and thereby enhances the stationary-state wealth inequality as shown in Eq.~(\ref{eq:sigma2eq}). 

\begin{figure*}
\includegraphics[width=2\columnwidth]{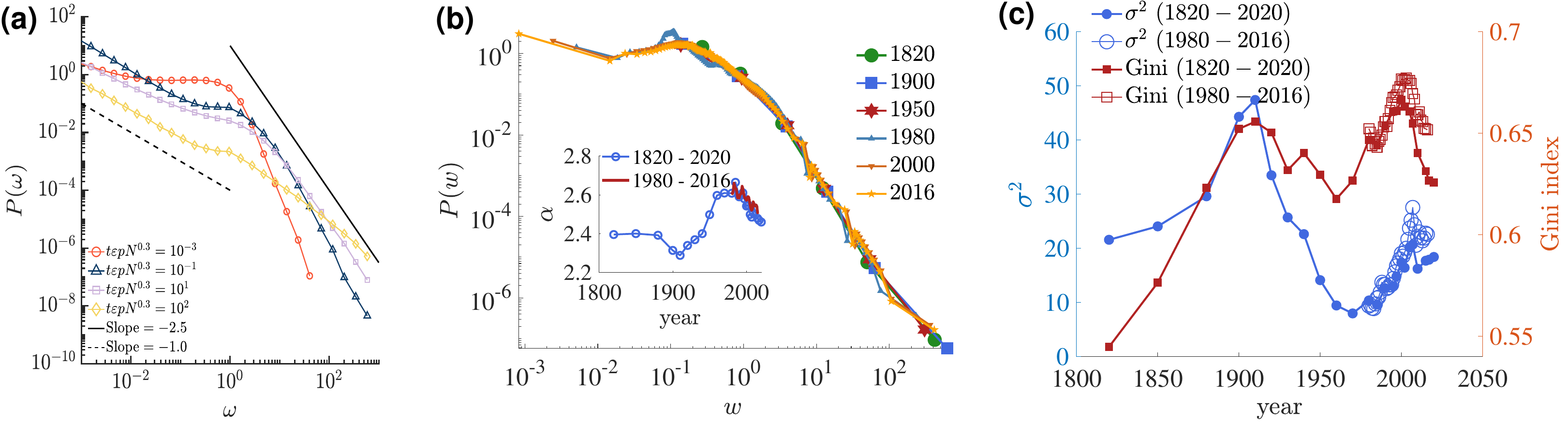}
\caption{Inequality in the model and the real world. (a) Wealth distributions in the generalized YS model with $p = 2.5 \times 10^{-6}, \ \varepsilon=0.05$ at different times on SF networks with $N=16000, \overline{k}=16$ and $\gamma=2.5$. The solid and dashed lines are guide lines with different slopes. 
(b) Income distributions in several years between 1820 and 2020 obtained from the data-sets in Refs.~\cite{chancel_piketty_2021} and ~\cite{chancel2022world}. Data have been rescaled so that the average is one, $\bar{\omega}=1$. Inset: The power-law exponent $\alpha$ in $P(\omega)\sim \omega^{-\alpha}$ fitted to the data-sets  for $\omega>10$.  
(c) Wealth variance $\sigma^2$ and the Gini index of the rescaled income in the period 1820 to 2020 (filled points) ~\cite{chancel_piketty_2021} and in the period 1980 to 2016 (open) ~\cite{chancel2022world}.  }
\label{fig:real}
\end{figure*}

Finally,  let us discuss the form of the wealth distribution on sparse networks. In the local condensate and the relaxation phase with small $p$ ($p\ll p_*$), there are a number of rich nodes as well as poor nodes. the poor nodes' wealth, mostly much smaller than the average $\omega\ll \overline{\omega}=1$, is found to be distributed by the power-law with exponent one, $P(\omega)\sim \omega^{-1}$,  the same as the stationary-state distribution in Eq.~\eqref{eq:pwp0}. However, the distribution of rich nodes' wealth is different, characterized by a larger power-law exponent than one as shown in Fig.~\ref{fig:real}(a).  As we have shown in the main text, the origin lies in the correlation of the wealth of a rich node with its degree 
\begin{equation}
\omega_{\rm rich}(k) \sim c \, k,
\label{eqS:omegarich}
\end{equation}
where the coefficient $c$ is close to one in the local condensation phase ($t_{\rm lc}\ll t\ll t_{\rm rel}$) and $c\simeq \varepsilon \, p\, t$ in the relaxation phase ($t_{\rm rel}\ll t \ll t_{\rm eq}$) as shown in Sec.~\ref{sec:intermediate}. Also, the probability of a node to be rich is given by 
\begin{equation}
\rho_{\rm rich}(k) \simeq 
\begin{cases}
{1\over k+1} &{\rm for} \ t_{\rm lc}\ll t\ll t_{\rm rel},\\
\rho_{\rm rich} &  {\rm for} \ t_{\rm rel}\ll t \ll t_{\rm eq}
\end{cases}
\label{eqS:rhorich}
\end{equation}
with $\rho_{\rm rich}$ given in Eq.~(\ref{eq:rho}).  Therefore the wealth distribution $P(\omega)$ for large $\omega$ is evaluated by using the degree distribution $P_{\rm deg}(k)$ as
\begin{align}
P(\omega) &\sim \sum_k P_{\rm deg}(k) \rho_{\rm rich}(k) \delta(\omega_{\rm rich}(k)-\omega)\cr
&\sim P_{\rm deg} \left(k\simeq {\omega\over c}\right)  \rho_{\rm rich} \left(k\simeq {\omega\over c}\right) \nonumber\\
&\sim \begin{cases}
\omega^{-\gamma-1} &{\rm for} \ t_{\rm lc}\ll t\ll t_{\rm rel},\\
\omega^{-\gamma} &  {\rm for} \ t_{\rm rel}\ll t \ll t_{\rm eq}.
\end{cases}
\label{eqS:pw}
\end{align}
This means that the power-law exponent of the wealth distribution $P(\omega)$ for large $\omega$ may be between $\gamma$ and $\gamma+1$, probably changing from $\gamma+1$ to $\gamma$ with time, in the intermediate-time regime [Fig.~\ref{fig:real}(a)]. Approaching the stationary state, the number of rich nodes decreases [Eq.~(\ref{eq:rho})] and therefore the fast-decaying right tail of $P(\omega)$ in Eq.~\eqref{eqS:pw} shrinks and eventually disappears in the stationary state. 

The real-world income distributions, which are obtained by compiling the world income data in Refs.~\cite{chancel_piketty_2021} and ~\cite{chancel2022world} as described in Appendix~\ref{seca:real},  decay slow for small income and then fast for large income with the latter characterized by a power-law of the exponent between 2 and  3. [Fig.~\ref{fig:real}(b)].  It is similar to the behavior of the wealth distribution in the intermediate-time regime; $P(\omega)\sim \omega^{-1}$ for small $\omega$ and Eq.~\eqref{eqS:pw} for large $\omega$. Different forms of the income distributions between the lower and upper class of the USA have been noted also in Ref.~\cite{doi:10.1098/rsta.2021.0162}. For comparison, we used the rescaled data so that the average income is equal to one. The income variance $\sigma^2$ is far from diverging with the population size, which would be the case with global wealth condensation; it increased from 20 to 50 in the period 1820 to 1900 and then decreased to less than 10 until around 1960, probably related to two world wars, and finally increased  reaching $\sigma^2\simeq 20$ in 2020 [Fig.~\ref{fig:real}(c)].

\section{Discussions}
\label{sec:discussion} 

In this study, we have characterized the scaling properties of wealth inequality, represented by wealth variance, in both its dynamics and steady-state. In the process, we have identified the critical value for redistribution ratio which is inversely proportional to the population size. Furthermore, the evolution of wealth inequality on a sparsely-connected population undergoes multiple phases, revealing the effects of network structure: If heterogeneously connected, a large population progresses into an inequality at a greater rate than a small one, while if homogeneously connected, the population size does not affect the speed. This is direct consequence of the correlation between wealth and node degree. Our study thus demonstrates that the relevance of the network structure to the wealth distribution in the non-stationary state should be considered in analyzing the real-world wealth inequality.

The influence of the dimensionality of the underlying network on the speed of wealth condensation deserves investigation. The studied model is a minimal one, and can be extended with more data-sets of the real-world inequality.

\begin{acknowledgments}
We thank Su-Chan Park for helpful discussion. This work was funded by  KIAS Individual Grants (CG079902 (D.-S.L) and CG084501 (H.G.L)) at Korea Institute for Advanced Study. We are grateful to the Center for Advanced Computation in KIAS for providing computing resources.
\end{acknowledgments}

\appendix 

\section{Boltzmann equation for $p=0$}
\label{seca:boltzmann}

When $p$ is small, in the long-time limit, the approximations that we take to obtain  the left or right tail of the wealth distributions in Eqs.~\eqref{eq:Plognormal} and \eqref{eq:Pgaussian}  are not working but one should refer to the Boltzmann equation~\cite{boghosian_kinetics_2014}
\begin{align*}
    &\frac{\partial P(\omega,t)}{\partial t} =  - \biggl[P(\omega,t) - \frac{1}{1+\varepsilon}P\Bigl(\frac{\omega}{1+\varepsilon},t\Bigr) \biggr] \\
 &    - \biggl[P(\omega,t) - \frac{1}{1-\varepsilon}P\Bigl(\frac{\omega}{1-\varepsilon},t\Bigr) \biggr] \\
    &+ \int_{0}^{\frac{\omega}{1+\varepsilon}}d\omega' \biggl[P(\omega-\varepsilon\omega',t) - \frac{1}{1+\varepsilon}P\Bigl(\frac{\omega}{1+\varepsilon},t\Bigr) \biggr] P(\omega',t) \\
    & + \int_{0}^{\frac{\omega}{1-\varepsilon}}d\omega' \biggl[P(\omega+\varepsilon\omega',t) - \frac{1}{1-\varepsilon}P\Bigl(\frac{\omega}{1-\varepsilon},t\Bigr) \biggr] P(\omega',t)   \numberthis
\end{align*}
under the random-agent approximation. Then, one can find the following solution by direct substitution:
\begin{align*}
    P(\omega,t) = \frac{C}{\omega (t + t_0)},    \numberthis
\end{align*}
where $t_0$ is a finite constant and $C$ is given by
\begin{align*}
    C = \frac{1}{\log{\bigl(\frac{1}{1-\varepsilon^2}\bigr)}}\simeq {1\over \epsilon^2} \numberthis
\end{align*}
with the last approximation valid for small $\epsilon$.

\section{Density of random walkers in the CRW: Derivation of Eq.~(\ref{eq:rho})}
\label{seca:CRW}

Following \cite{catanzaro_diffusion_2005}, we can see that the number of random walkers, $n_i=0$ or $1$, on node $i$ at time $t+dt$ in the CRW model is given by
\begin{align*}
    n_i(t+dt) = n_i(t)\eta + (1 - n_i(t))\xi,    \numberthis
\end{align*}
where $\eta$ is $0$ with probability $ \lambda dt \sum_j a_{ij}(1-n_j)+\lambda dt\sum_j a_{ij}n_j = \lambda\,  dt \, k_i$ and $1$ with probability $1-\lambda \, dt \,k_i$, and $\xi$ is $1$ with probability $\lambda \, dt\sum_j a_{ij}n_j$ and $0$ with probability $1-\lambda \,dt\sum_j a_{ij}n_j$. Here, $\eta=0$ represents the disappearance of the random-walker at node $i$ because of a jump to a nearest node (with probability $\lambda dt\sum_j a_{ij}(1-n_j)$), or coalescence to a neighbor node (with probability $\lambda dt\sum_j a_{ij}n_j$). On the other hand, $\xi=1$ represents the arrival of a walker at node $i$ from a neighbor node (with probability $\lambda dt\sum_j a_{ij}n_j$). Taking the ensemble average for a given $n_i(t)$, we have 
\begin{align*}
    \langle n_i(t+dt) \rangle &= n_i(t)\big(1-\lambda\, dt\, k_i\big) \\
    &+\big(1-n_i(t)\big) \lambda dt \sum_j a_{ij}n_j(t),    \numberthis    
\end{align*}
which leads to
\begin{align*}
    \frac{d\rho_i}{dt} = - \lambda k_i \rho_i + \lambda \big(1-\rho_i\big)\sum_j a_{ij}\rho_j , \numberthis
    \label{eq:drhoi}
\end{align*}
where the density of random walker is denoted by $\rho_i=\langle n_i \rangle$ and the independence of $\rho_i$ and $\rho_j$ for $i\neq j$ is assumed. Assuming that $\rho_i$ at node $i$ is a function of its degree $k_i$ and equivalently $\rho_i=\rho_j$ if $k_i=k_j$, we can rewrite Eq.~(\ref{eq:drhoi}) as 
\begin{align*}
    \frac{d\rho_k}{dt} = - \lambda k \rho_k + \lambda k\big(1-\rho_k\big)\sum_{k'} \frac{k' P_{\rm deg}(k')}{\overline{k}}\rho_{k'},      \numberthis
    \label{eqS:drhok}
\end{align*}
where $\rho_k \equiv \frac{1}{N}\sum_{i=1}^{N}\delta_{k_i,k} \ \rho_i$ and $P_{\rm deg}(k)$ is the degree distribution of the underlying network. Note that $\rho=\sum_k P_{\rm deg}(k)\rho_k$. Let us also introduce $\tilde{\rho} \equiv \sum_k \frac{k P_{\rm deg}(k)}{\overline{k}}\rho_{k}$. Multiplying Eq.~(\ref{eqS:drhok}) by $P_{\rm deg}(k)$ and summing over $k$, we find
\begin{align*}
    \frac{d\rho}{dt}=-\lambda \overline{k} \tilde{\rho} + \lambda \overline{k} \tilde{\rho}\big(1-\tilde{\rho}\big) = -\lambda \overline{k} \tilde{\rho}^2.    \numberthis
    \label{eqS:drhodt}
\end{align*}
In the long-time limit, we expect $\frac{d\rho_k}{dt}$ is much smaller than $\rho_k$ or $\tilde{\rho}$ in Eq.~(\ref{eqS:drhok}), so we assume that the right-hand-side of Eq.~(\ref{eqS:drhok}) is zero, to obtain
\begin{align*}
    \rho_k \simeq \frac{\tilde{\rho}}{1+\tilde{\rho}},    \numberthis
\end{align*}
which means that $\rho_k$ is independent of $k$. This also implies that 
\begin{align*}
   \rho=\sum_k P_{\rm deg}(k)\rho_k=\frac{\tilde{\rho}}{1+\tilde{\rho}} \quad  \text{or}  \quad  \tilde{\rho}=\frac{\rho}{1-\rho}.       \numberthis
    \label{eqS:rhorhotilde}
\end{align*}
Inserting Eq. (\ref{eqS:rhorhotilde}) into Eq.~(\ref{eqS:drhodt}), we have
\begin{align*}
    \frac{d\rho}{dt}=- \lambda \overline{k} \bigg(\frac{\rho}{1-\rho} \bigg)^2 \simeq - \lambda \overline{k} \rho^2,      \numberthis
    \label{eqS:drhodtapprox}
\end{align*}
where the last relation holds for $\rho \ll 1$ which will be valid in the long-time limit. Therefore, from Eq. (\ref{eqS:drhodtapprox}), we have 
\begin{align*}
    \rho \simeq \frac{1}{\lambda \overline{k} t},     \numberthis
\end{align*}
which gives  $\rho_{\rm rich} \sim \rho \sim \frac{1}{\varepsilon p \overline{k} t}$ with $\lambda_{gYS}=\varepsilon p$ as in Eq.~(\ref{eq:rho}) in the main text. It takes $t_{\rm hop}\sim {1\over \varepsilon p}$ for the local wealth at a rich node $i$ to be fully redistributed to its poor neighbors and itself, which is understood by considering  ${\varepsilon p\over \overline{k}} k_i (\omega_i-1) t_{\rm hop} \sim  \omega_i-1$ and essentially determines the rate $\lambda_{\rm gYS}\simeq {1\over t_{\rm hop}}\sim \varepsilon p$ of random hopping of local wealth.

\section{Income distribution in the real world in 1820 - 2020}
\label{seca:real}

We use two data-sets of the world income distribution, one in Ref.~\cite{chancel_piketty_2021} covering  a few selected years in the period 1820 to 2020 and the other in Ref.~\cite{chancel2022world} providing annual data from 1980 to 2016. Both data-sets provide for each given year  the average income $W(x)$ of the people who  belong to the top $x$ \% of the whole population in the world regarding their income with $x = 0.01, 0.1, 1, 10, 50$, and $100$. We can consider 6 distinct groups of people according to their income: bottom 50\%, top 50\% to top 10\%, top 10\% to top 1\%, top 1\% to top 0.1\%, top 0.1\% to top 0.01\%, and the top 0.01\%. The average incomes of the people in these 6 distinct groups and their portions are then given by
\begin{equation}
\begin{cases}
W_1 = W(100) - W(50), \ f_1 = 0.5,\\
W_2 = W(50)- W(10), \ f_2 = 0.4,\\
W_3 = W(10) - W(1), \ f_3 = 0.09,\\
W_4 = W(1) - W(0.1), \ f_4 = 0.009,\\
W_5 = W(0.1) - W(0.01), \ f_5 = 0.0009,\\
W_6 = W(0.01), \ f_6 = 0.0001.
\end{cases}
\end{equation}
Notice that $\sum_{\ell=1}^6 f_i=1$. We can use these values to compute the mean wealth 
\begin{equation}
\overline{W} = \sum_{\ell=1}^6 W_\ell f_\ell.
\end{equation}

 To compare the properties of these real-world incomes with the wealth considered in the main text,  we consider the rescaled income 
\begin{equation}
\omega_\ell = {W_\ell\over \overline{W}}
\end{equation}
for $\ell = 1,2,3,4,5,6$. 
The income variance is then evaluated as
\begin{equation}
\sigma^2 = \sum_{\ell=1}^6 \omega_\ell^2 f_\ell -1 
\end{equation}
and the (rescaled) income distribution $P(\omega)$ is obtained by 
\begin{equation}
p(\omega_\ell) = {f_\ell\over \Delta \omega_\ell},
\end{equation}
where the bin sizes are evaluated as $\Delta \omega_\ell = \sqrt{\omega_{\ell} \omega_{\ell+1}} - \sqrt{\omega_{\ell}\omega_{\ell-1}}$ with $\omega_{0} = {\omega_1^2\over \sqrt{\omega_1\omega_2}}$ and $\omega_7 =  {\omega_6^2\over \sqrt{\omega_5\omega_6}}$. We also compute the Gini index by 
\begin{equation}
G = {1\over 2} \sum_{\ell_1 = 1}^6 \sum_{\ell_2=1}^6 |\omega_{\ell_1} - \omega_{\ell_2}| f_{\ell_1} f_{\ell_2},
\end{equation}
where we used $\overline{\omega}=1$. Note that the wealth variance $\sigma^2$ is represented by $\sigma^2 = (1/2)\sum_{\ell_1, \ell_2} (\omega_{\ell_1} - \omega_{\ell_2})^2 f_{\ell_1} f_{\ell_2}$.


\bibliography{main.bib}

\end{document}